\input harvmac
\input amssym.def
\input amssym.tex
\noblackbox
\newif\ifdraft

\catcode`\@=11
\newif\iffrontpage
\newif\ifxxx
\xxxtrue

\newif\ifad
\adtrue
\adfalse

\parindent0pt

\def\a{\alpha}
\def\b{\beta}

\def\s{\sigma}

\def\p{\partial}


\parindent0pt

\def\{{\lbrace}
\def\}{\rbrace}

\def\a{\alpha}
\def\b{\beta}

\def\s{\sigma}

\def\p{\partial}



\def\{{\lbrace}
\def\}{\rbrace}

\def\a{\alpha}
\def\b{\beta}

\def\s{\sigma}

\def\p{\partial}

\def\frac#1#2{{\scriptstyle{#1}\over\scriptstyle{#2}}}

\def\abstract#1{
\vskip.5in\vfil\centerline
{\bf Abstract}\penalty1000
{{\smallskip\ifx\answ\bigans\leftskip 2pc \rightskip 2pc
\else\leftskip 5pc \rightskip 5pc\fi
\noindent\abstractfont \baselineskip=12pt
{#1} \smallskip}}
\penalty-1000}

\lref\ChuPB{
  C.~S.~Chu and D.~Giataganas,
  ``1/4 BPS Wilson Loop in beta-deformed Theories,''
  JHEP {\bf 0710}, 108 (2007)
  [arXiv:0708.0797 [hep-th]].
}

\lref\KhozeND{
  V.~V.~Khoze,
  ``Amplitudes in the beta-deformed conformal Yang-Mills,''
  JHEP {\bf 0602}, 040 (2006)
  [arXiv:hep-th/0512194].
}

\lref\ElmettiUP{
  F.~Elmetti, A.~Mauri, S.~Penati, A.~Santambrogio and D.~Zanon,
  ``Real versus complex beta-deformation of the N=4 planar super Yang-Mills
  theory,''
  JHEP {\bf 0710}, 102 (2007)
  [arXiv:0705.1483 [hep-th]].
}

\lref\MauriPA{
  A.~Mauri, S.~Penati, A.~Santambrogio and D.~Zanon,
  ``Exact results in planar N = 1 superconformal Yang-Mills theory,''
  JHEP {\bf 0511}, 024 (2005)
  [arXiv:hep-th/0507282].
}

\lref\DrummondBM{
  J.~M.~Drummond, J.~Henn, G.~P.~Korchemsky and E.~Sokatchev,
  ``The hexagon Wilson loop and the BDS ansatz for the six-gluon amplitude,''
  arXiv:0712.4138 [hep-th].

}

\lref\DrummondCF{
  J.~M.~Drummond, J.~Henn, G.~P.~Korchemsky and E.~Sokatchev,
  ``On planar gluon amplitudes/Wilson loops duality,''
  arXiv:0709.2368 [hep-th].
}

\lref\DrummondAUA{
  J.~M.~Drummond, G.~P.~Korchemsky and E.~Sokatchev,
  ``Conformal properties of four-gluon planar amplitudes and Wilson loops,''
  arXiv:0707.0243 [hep-th].
}

\lref\AKS{
S.~Ananth, S.~Kovacs and H.~Shimada,
``Proof of all-order finiteness for planar beta-deformed Yang-Mills,''
JHEP {\bf 0701}, 046 (2007)
[arXiv:hep-th/0609149].}

\lref\Filk{
T.~Filk,
``Divergencies in a field theory on quantum space,''
Phys.\ Lett.\  B {\bf 376}, 53 (1996).}

\lref\MVS{
S.~Minwalla, M.~Van Raamsdonk and N.~Seiberg,
``Noncommutative perturbative dynamics,''
JHEP {\bf 0002}, 020 (2000)
[arXiv:hep-th/9912072].}

\lref\LS{
R.~G.~Leigh and M.~J.~Strassler,
``Exactly Marginal Operators And Duality In Four-Dimensional N=1
Supersymmetric Gauge Theory,''
Nucl.\ Phys.\  B {\bf 447}, 95 (1995)
[arXiv:hep-th/9503121].}

\lref\AdSreview{
O.~Aharony, S.~S.~Gubser, J.~M.~Maldacena, H.~Ooguri and Y.~Oz,
``Large N field theories, string theory and gravity,''
Phys.\ Rept.\  {\bf 323}, 183 (2000)
[arXiv:hep-th/9905111].}

\lref\Maldacena{
J.~M.~Maldacena,
``The large N limit of superconformal field theories and supergravity,''
Adv.\ Theor.\ Math.\ Phys.\  {\bf 2}, 231 (1998)
[Int.\ J.\ Theor.\ Phys.\  {\bf 38}, 1113 (1999)]
[arXiv:hep-th/9711200].}

\lref\AMI{
L.~F.~Alday and J.~M.~Maldacena,
``Gluon scattering amplitudes at strong coupling,''
JHEP {\bf 0706}, 064 (2007)
[arXiv:0705.0303 [hep-th]].}

\lref\LM{
O.~Lunin and J.~M.~Maldacena,
``Deforming field theories with U(1) x U(1) global symmetry and their
gravity duals,''
JHEP {\bf 0505}, 033 (2005)
[arXiv:hep-th/0502086].}

\lref\BDS{
Z.~Bern, L.~J.~Dixon and V.~A.~Smirnov,
``Iteration of planar amplitudes in maximally supersymmetric Yang-Mills
theory at three loops and beyond,''
Phys.\ Rev.\  D {\bf 72}, 085001 (2005)
[arXiv:hep-th/0505205].}

\lref\KRTT{
M.~Kruczenski, R.~Roiban, A.~Tirziu and A.~A.~Tseytlin,
``Strong-coupling expansion of cusp anomaly and gluon amplitudes from quantum
open strings in $AdS_5\times S^5$,''
Nucl.\ Phys.\  B {\bf 791}, 93 (2008)
[arXiv:0707.4254 [hep-th]].}

\lref\MT{
R.~R.~Metsaev and A.~A.~Tseytlin,
``Type IIB superstring action in $AdS_5\times S^5$ background,''
Nucl.\ Phys.\  B {\bf 533}, 109 (1998)
[arXiv:hep-th/9805028].}

\lref\AAF{
L.~F.~Alday, G.~Arutyunov and S.~Frolov,
``Green-Schwarz strings in TsT-transformed backgrounds,''
JHEP {\bf 0606}, 018 (2006)
[arXiv:hep-th/0512253].}

\lref\FGT{
S.~Forste, D.~Ghoshal and S.~Theisen,
``Stringy corrections to the Wilson loop in N = 4 super Yang-Mills  theory,''
JHEP {\bf 9908}, 013 (1999)
[arXiv:hep-th/9903042].}

\lref\DrummondAU{
  J.~M.~Drummond, J.~Henn, G.~P.~Korchemsky and E.~Sokatchev,
  ``Conformal Ward identities for Wilson loops and a test of the duality with
  gluon amplitudes,''
  arXiv:0712.1223 [hep-th].
}

\lref\FrolovDJ{
  S.~Frolov,
  ``Lax pair for strings in Lunin-Maldacena background,''
  JHEP {\bf 0505}, 069 (2005)
  [arXiv:hep-th/0503201].}

\lref\AnanthPX{
  S.~Ananth, S.~Kovacs and H.~Shimada,
  Nucl.\ Phys.\  B {\bf 783}, 227 (2007)
  [arXiv:hep-th/0702020].
}

\lref\BrandhuberYX{
  A.~Brandhuber, P.~Heslop and G.~Travaglini,
  ``MHV Amplitudes in N=4 Super Yang-Mills and Wilson Loops,''
  arXiv:0707.1153 [hep-th].
}

\lref\McGreevyKT{
  J.~McGreevy and A.~Sever,
  ``Quark scattering amplitudes at strong coupling,''
  arXiv:0710.0393 [hep-th].
}

\lref\KomargodskiER{
  Z.~Komargodski and S.~S.~Razamat,
  ``Planar quark scattering at strong coupling and universality,''
  arXiv:0707.4367 [hep-th].
}


\Title{\vbox{ 
\rightline{\vbox{\baselineskip12pt
\hbox{$\scriptscriptstyle{.}$}}}}}
{Gluon Scattering in Deformed ${\cal N}=4$ SYM
\footnote{$^{\scriptscriptstyle*}$}{\sevenrm Partially
supported by GIF, the German-Israeli Foundation for Scientific
Research and by the German Israeli Project Cooperation (DIP H52)}}
\vskip 0.3cm
\centerline{ Yaron Oz$^a$,  Stefan Theisen$^b$ and Shimon Yankielowicz$^a$ }
\vskip 0.6cm
\centerline{$^a$ \it Sackler School of Physics, Tel-Aviv University, Israel}
\vskip.2cm
\centerline{$^b$ \it Max-Planck-Institut f\"ur Gravitationsphysik,
Albert-Einstein-Institut, 14476 Golm, Germany}

\bigskip
\vskip 1cm

\centerline{Abstract}

\vskip 0.5cm
We consider gluon and gluino scattering amplitudes in large $N$ $\b$-deformed ${\cal N}=4$ SYM with
real $\b$.
A direct inspection of the planar diagrams shows that the scattering amplitudes to all
orders in perturbation theory are the same as in the undeformed ${\cal N}=4$ SYM theory.
Using the dual $\sigma$-model description, we find the same equality at strong coupling to all
orders in the $\sigma$-model loop expansion.
Finally, we show that the same analysis holds for
gluon scattering amplitudes in a three-parameter deformation of 
planar ${\cal N}=4$ SYM that breaks all the supersymmetry.

\Date{\vbox{\hbox{\sl {December 2007}}
}}
\goodbreak

\parskip=4pt plus 15pt minus 1pt
\baselineskip=15pt plus 2pt minus 1pt

\noblackbox

\newsec{Introduction}

Recently, a possible intriguing duality has been revealed in ${\cal N}=4$ SYM theory at the
planar limit.
The planar MHV gluon scattering amplitudes ${\cal A}$
in the theory seem to have a dual description in terms of
the expectation value of a Wilson loop, whose contour $C$
consists of  light-like segments, which are
proportional to the light-like momenta $k_i$ of the external gluon
\eqn\wilson{
{\cal A} = {1 \over N}\langle{\rm tr} P \exp (i \oint_C A_{\mu}(x)dx^{\mu})\rangle + O(1/N^2)}

A prescription, along these lines, for the computation of the planar gluon scattering amplitudes in
${\cal N}=4$ SYM at
strong coupling has been proposed in \AMI.
The kinematic factor of the color ordered planar
scattering amplitude ${\cal A}$ was mapped to the computation of the
Wilson loop, which
using the gauge/gravity correspondence  \Maldacena,\AdSreview,
translates to the construction of a minimal area string worldsheet
ending on the loop.
The exponential of this (regularized) area represents
the gluon scattering amplitude
\eqn\area{
{\cal A} \sim \exp (-{\sqrt{\lambda}\over 2\pi} Area)
}
where $\lambda = g^2N$.
The four-gluon scattering amplitude at strong coupling  $\lambda \gg 1$ was computed
in this way in \AMI, and was shown to be in agreement with the
general structure conjectured by Bern, Dixon and Smirnov in \BDS.
Evidence for the duality at weak coupling has been given recently, at one-loop in \DrummondAUA\BrandhuberYX, at two-loop in \DrummondCF\DrummondAU, and for its possible breakdown for six
gluons in \DrummondBM.
A study of quark scattering amplitudes has been done in  \KomargodskiER\McGreevyKT.

A natural question to ask is whether there are other theories with less supersymmetry
where this duality can hold.
In this note we will
consider gluon and gluino scattering amplitudes and the corresponding light-like Wilson loops
in large $N$ $\b$-deformed ${\cal N}=4$ SYM with
real $\b$ (for a discussion of Wilson loops in the $\beta$-deformed theory see \ChuPB).
This deformation breaks the ${\cal N}=4$ supersymmetry to ${\cal N}=1$.
We will further analyse a $\gamma$-deformation of ${\cal N}=4$ SYM that breaks supersymmetry completely.

The note is organized as follows. In section 2 we will analyze the planar gluon scattering amplitudes and Wilson
loop in the weak coupling regime of the theory.
We will argue, based on the observation that the $\b$-deformed theory can be written as a non-commutative
deformation of the ${\cal N}=4$ theory, that to all orders in perturbation
theory they have exactly the same values as in ${\cal N}=4$ SYM.
In section 3 we will analyze the strong coupling regime and use the $\sigma$-model description of the
$\b$-deformed theory and show that the same holds to all orders in the  $\sigma$-model loop expansion.
In section 4 we will consider planar gluon scattering amplitudes and the corresponding Wilson loop expectation value
in a three-parameter deformation of ${\cal N}=4$ SYM that breaks all the supersymmetry.
We will see that the same analysis done for the $\b$-deformed theory applies also here.

In this note we assume that the deformation parameter $\b$ is real.
One may also consider the case where $\b$ is complex. The resulting deformed theory
was argued to be conformal \LS (for a recent discussion see \ElmettiUP).
On the perturbative field theory side, inspection of the planar diagrams shows that unlike
the case with real $\b$, here there is a dependence of the scattering amplitudes on the deformation parameter. However, it is straightforward to check that the dependence on $\beta$ does not appear
up to the two-loop order both in the gluon scattering amplitudes and in the Wilson loop. 
In the strong coupling regime, the dual supergravity background is of a warped type. It
was obtained by employing an S-duality transformation \LM, and we lack a simple $\sigma$-model
description.

\newsec {Weak coupling analysis}

We will consider the $\b$-deformation of ${\cal N}=4$ SYM with $SU(N)$ gauge group at  large $N$.
In the following we will  take the parameter $\b$ to be real.
The deformed theory has ${\cal N}=1$ supersymmetry and is conformally invariant \LS\MauriPA\AKS.
The field content of the $\b$-deformed theory is identical to that of
${\cal N}=4$ SYM, i.e.
it consists, in ${\cal N}=1$ language, of a vector superfield
$V\sim(A_\mu,\lambda_\alpha)$
and three chiral superfields $\Phi^I\sim(\phi^I,\psi^I_\alpha),\,I=1,2,3$,
all transforming in the adjoint representation of
the gauge group. The $SU(4)$ ${\cal R}$-symmetry of  ${\cal N}=4$ SYM is broken by the deformation to
$U(1)_{\cal R}\times (U(1)\times U(1))_{\rm flavor}$.
The vector multiplet fields are neutral under the flavor
symmetry, while the field three chiral multiplets carry the charges
$(Q^1,Q^2)=(0,-1),(1,1)$ and $(-1,0)$, respectively.

Written in terms of ${\cal N}=1$ superfields, the action of the $\b$-deformed
theory is that of three adjoint chiral multiplets coupled to a
vector multiplet and a superpotential
\eqn\superpotential{
W=g\, {\rm tr}(q\, \Phi^1\Phi^2\Phi^3-\bar q\,\Phi^1\Phi^3\Phi^2)
=g_{abc}\Phi_1^a\Phi_2^b\Phi_3^c
}
where $g$ is the gauge coupling, $q=e^{i\pi\b}$ ($|q|=1$ for real $\b$)
and
\eqn\gabc{
g_{abc}=g(q\, c_{abc}-q^{-1}\, c_{acb})\,,\qquad c_{abc}={\rm tr}(T^a T^b T^c)}

The superpotential can be written
in the form  $d$-type and a $f$-type coupling as in \LS,
\eqn\wabc{
W= 2 g \, (\cos[\pi\b] f_{abc} + i \sin[\pi\b] d_{abc}) \Phi_1^a\Phi_2^b\Phi_3^c
}

where
$f_{abc}={\rm tr}(T^a [T^b, T^c]), d_{abc}={\rm tr}(T^a \{T^b, T^c\})$.
The latter term in $W$ vanishes for
$\b=0$ where one recovers the ${\cal N}=4$ theory.
Note that the ${\cal N}=4$ SYM action and its $\b$-deformation have the same propagators and chiral-vector vertices and differ
only in the chiral vertex.

The Coulomb branch of the $\b$-deformed theory is a solution to the F-term equations
\eqn\coulomb{
\Phi^1\Phi^2 = q\Phi^2\Phi^1\,,\qquad  \Phi^2\Phi^3 = q\Phi^3\Phi^2\,,\qquad  \Phi^3\Phi^1 = q\Phi^1\Phi^3
}

A useful observation is that the $\b$-deformed action can be written
using  a noncommutative $*$-product between the matter fields
defined as  \LM
\eqn\defstar{
f*g=e^{i\pi\beta(Q_f^1 Q_g^2-Q_f^2 Q_g^1)}fg}
where $fg$ is the ordinary product and $(Q^1,Q^2)$ are the $(U(1)\times U(1))_{\rm flavor}$ charges of the matter fields
(the vector multiplet is neutral under the flavor group).
This has been employed in  \AKS\ with
${\cal N}=4$ light-cone superspace to prove the
finiteness of the $\b$-deformed theory at large $N$.

By a direct inspection of Feynman diagrams as in \Filk\MVS\KhozeND,
it is straightforward to see that the only modification of {\it planar}
diagrams compared to ${\cal N}=4$ SYM, is the multiplication by an overall phase, which depends only
on the flavor charges of the fields on the external lines. In particular if
all the fields on the external lines are gluons and gluinos, the phase is 1 and
the scattering amplitudes are same as in the
undeformed   ${\cal N}=4$ SYM theory. This holds to all orders in perturbation theory for any
$n$-point function.
The same argument can be employed in the computation of the expectation value
of the Wilson loop in the planar limit of the $\b$-deformed theory and shows that it agrees with the ${\cal N}=4$ SYM result to all
orders in perturbation theory.


\newsec{Strong coupling analysis}

We have argued in the previous section that at weak coupling the $\b$ deformation is invisible in
the planar gluon and gluino scattering amplitudes to all orders in perturbation theory.
In this section we will consider the
$\b$-independence of the scattering amplitudes at strong coupling.
We will use the $\sigma$-model description of the dual supergravity background and show that
that it holds at strong coupling to all
orders in the $\sigma$-model loop expansion.

As we discussed above, the prescription for the computation of the planar gluon scattering amplitudes in
${\cal N}=4$ SYM at
strong coupling  was mapped using the gauge/gravity correspondence to
the construction of a minimal area string worldsheet
ending on the loop \AMI.

There are two ingredients in the computation that we will need to pay special
attention to. The first one is that in order to map
the scattering amplitude computation to the Wilson loop one at strong coupling, a
T-duality on all four directions along the boundary of $AdS_5$ has been employed.
The loop is the boundary of the T-dualized open string world sheet in
$AdS$ space, corresponding to the IR region in field theory.
The second one is that the area of the minimal surface is infinite and needs regularization.
The infinity corresponds to the infrared divergences
in the field theory.
One type of regularizaton introduced in \AMI, was in the form of a D3-brane placed at a large
value of the $AdS_5$ radial coordinate $Z_{IR}$, and at an arbitrary point in the internal $S^5$ space.
In the field theory language it translates to going
along the Coulomb branch.
The Coulomb branch of the $\b$-deformed theory \coulomb\ differs from that of
${\cal N}=4$ SYM. Therefore, if we wish to employ a similar regularizaton procedure, we will
have to place the D3-brane in a particular locus in the internal part
of the space.

{\bf The dual supergravity description}

The dual gravity background
of the $\b$-deformed theory at large $N$
 was constructed in
\LM. The metric for real $\beta$
is of the type $AdS_5 \times {\cal M}_5$, where ${\cal M}_5$ is a deformation of $S^5$.
With  the $S^5$  parameterized by
three angular coordinates $\phi_i$ and three radial coordinates
$\mu_i$ satisfying $\mu_1^2+\mu_2^2+\mu_3^2=1$, the metric on  ${\cal M}_5$ takes the form
\eqn\metric{
ds^2_{{\cal M}_5} = \sum_{i=1}^3 (d\mu_i^2 + G \mu_i^2 d \phi_i^2) + \b R^2 G \mu_1^2\mu_2^2\mu_3^2(\sum_{i=1}^3d\phi_i)^2
}
where $G^{-1} = 1 + \b R^2  (\mu_1^2\mu_2^2+\mu_2^2\mu_3^2+\mu_1^2\mu_3^2)$ and $R$ is the $S^5$ radius.
In addition to this deformation of the $S^5$-part of the metric and the
field strength $F^{(5)}$, the deformed background has all
bosonic Type IIb supergravity fields turned on. However, the modification of all
the
background fields depends completely on the compact part ${\cal M}_5$.
This means, in particular, that the T-duality transformation in the $AdS_5$ part, that is used in order to map
the scattering amplitude computation to a Wilson loop one is still
valid in the $\b$-deformed geometry.

The deformation of $S^5$ to ${\cal M}_5$ is obtained
by  $TsT$ transformation:
a combination of a T-duality
along one of the circles, followed by a shift a along the second circle and
a second T-duality along the first circle.
The flat directions corresponding to the field theory Coulomb branch are characterized
by the points in ${\cal M}_5$
where two of the $\mu_i$'s vanish. Indeed, it is straightforward to see that
a D3-brane that is moved in the radial direction of $AdS_5$ and is located at these
points in the internal space experiences zero force.
Thus, the analog of the regularization procedure of \AMI, requires a location of the D3-brane
at this locus of points.
Obviously, this is expected since the Coulomb branch of ${\cal N}=4$ SYM differs from
that of its deformation.
Note also that along this locus
the dependence of all the background fields on $\beta$ disappears.
We should stress, however, that we are computing the gluon scattering amplitude and the corresponding expectation value
of the Wilson loop at the origin of the Coulomb branch. Moving the D3-brane along the Coulomb branch is only one
type of regularization (dimensional regularization can be another method) and the fact that it can be located only at a
certain locus in ${\cal M}_5$
is not a restriction on the validity of the results, but only on the use of this type of regularization.

The computation of the minimal surface in \AMI was done at fixed position in the
internal space and the solution does not depend on the
coordinates of $S^5$.
This minimal surface solution is also a solution in the $\b$-deformed background and
is independent
of the deformation parameter.
Thus, the result of \AMI carries over to the $\b$-deformed case, and one gets exactly
the same four-gluon scattering amplitude as that of ${\cal N}=4$ SYM at strong coupling.
Similarly, the $n$-gluon scattering amplitudes are independent of $\b$ at strong coupling and are
the same as those of ${\cal N}=4$ SYM.

${1\over\sqrt{\lambda}}$ corrections to the result of \AMI\ are due to
fluctuations around the classical minimal surface solution.
To leading order in $N$ they are $\sigma$-model loops on the world sheet
with the disc topology.
The $\sigma$-model corrections to \AMI have been considered in
\KRTT. This analysis uses the explicit form of the Green-Schwarz action
in the $AdS_5\times S^5$ background \MT\ to compute the fluctuation spectrum
of the world-sheet fields.

As we discussed above, the $\b$-deformed background is independent of $\b$
at those points where
two of the $\mu_i$'s vanish.
Let us fix the internal coordinate of the minimal surface solution to one
of those points and consider the fluctuation spectrum around this solution.
The procedure is the same as in \FGT and
uses a covariant background field expansion and Riemann
normal coordinates. For the bosonic fluctuations along the tangent space of
$AdS_5$ there are no changes compared to the undeformed case.
The fluctuations tangential to the internal space is, in general,
affected. However, due to the fact that the classical solution satisfies
$\phi_i=\mu_i={\rm const.}$ and furthermore, say, $\mu_1=\mu_2=0$,
the fluctuation determinant will again be as in the undeformed
background geometry.\foot{In general there are term of the form
$\sqrt{h}h^{ij}R_{\mu\nu\rho\sigma}\p_i X^\mu_{cl}\p_j X^\rho_{cl}\xi^\nu\xi^\sigma$
where $h_{ij}$ is the world-sheet metric and $\xi^\mu$ the bosonic fluctuations.
This term vanishes for $X_{cl}={\rm const.}$}
The same holds for the fermions. Thus, up to the quadratic order
the action for the fermionic fluctuations is the same as in the undeformed case.
This shows that the $1/\sqrt{\lambda}$ corrections to the strong coupling result
are unmodified by the $\b$-deformation, at least if we place the minimal surface which ends
on the light-like Wilson loop at a point where two of the $\mu_i$'s vanish.

In the following we will show using the Green-Schwarz $\sigma$-model on the $\b$-deformed
background that the requirement of placing the
solution at those special points corresponding to the Coulomb branch is not needed.
Moreover, we will argue that the
results are the same as in the undeformed ${\cal N}=4$ SYM theory
at strong coupling to all
orders in the $\sigma$-model loop expansion.

{\bf The $\sigma$-model description}

A Green-Schwarz $\sigma$-model describing a closed string propagating
on the $\b$-deformed
background was constructed in \AAF.
The authors of \AAF\ have shown that the deformation of the
Green-Schwarz  $\sigma$-model action can be replaced by a modification of the periodicity
properties of the world-sheet fields. More specifically, they showed in general that the
angular coordinates $\phi_i$ in the compact space, which were involved in the $TsT$-transformation
are no longer periodic in $\s\to\s+2\pi$.
The difference $\phi(\s+2\pi)-\phi(\s)~({\rm mod}\,2\pi)$ is proportional to the deformation
parameter and to the $U(1)$ Noether charges associated to the two isometries
$\phi_i\to\phi_i+{\rm const}$.
In the following we will use the notations of \AAF.

For a purely bosonic background the action reads ($F_\alpha$ and $B_\alpha$
vanish)
\eqn\action{
{\cal L}= \tr[\gamma^{\a\b}\p_\a G G^{-1}\p_\b G G^{-1}]
}
where
$G=\pmatrix{g_a&0\cr0&g_s}$, and  $g_a\in SU(2,2)$ and $g_s\in SU(4)$ provide a parametrization
of $AdS_5$ and $S^5$.
For us the relevant piece is $g_s$, which gives
\eqn\act{
{\cal L}\sim \gamma^{\a\b}\sum_{i=1}^3(\mu_i^2\p_\a\phi_i\p_\b\phi_i+\p_\a\mu_i\p_\b\mu_i)
}
This implies that the  Noether currents,
whose charge are responsible for the
non-periodicity of the fields depend only on $\p^\alpha \phi$  and not on $\phi_i$ ($U^0_{\b,i}=V^0_{\b,i}=0$)
\eqn\currents{
J^\alpha_i \sim \partial^{\alpha}\phi_i
}
For constant $\phi_i$, as is
the case in the internal part of the minimal area solution, this vanishes.
According to \AAF\ this also means that the
fermionic fields are still periodic. Combining these arguments we conclude that
the analysis of the spectrum of fluctuations is unmodified by the deformation.

In fact, the calculation of the loop corrections to the light-like Wilson loop
in the $\sigma$-models of $AdS_5\times S^5$ and its TsT deformation will give the same result to all orders.
The reason being, that the classical solution dictates the periodicity of the fields and those
are $\b$-independent for the case at hand. Therefor, the corrections to any loop order
in both $\sigma$-models are identical.

\newsec {Nonsupersymmetric deformation}

In this section we will consider planar gluon scattering amplitudes and the corresponding Wilson loop expectation value
in a three-parameter deformation of ${\cal N}=4$ SYM that breaks all the supersymmetry. The resulting nonsupersymmetric deformed theory is scale invariant
in the planar limit \AnanthPX.
We denote the three real deformation parameters by $\gamma_i, i=1,2,3$ and refer to the deformed theory as $\gamma$-deformed.
When all the deformation parameters are equal $\gamma_i=\beta, i=1,2,3$, we obtain the $\b$-deformed theory that we discussed previously.
The $\gamma$-deformed theory  has the same field content as ${\cal N}=4$ SYM, with a modification of the ${\cal N}=4$ SYM
Yukawa and scalar quartic couplings by phase factors.
These phases break the $SU(4)$ ${\cal R}$-symmetry to its Cartan subgroup, now being  a flavor symmetry $U(1)^3_{\rm flavor}$ of the deformed
theory.
The six scalars and all the Weyl fermions including the gaugino are charged under  $U(1)^3_{\rm flavor}$. The gauge field is not charged under the
$U(1)^3_{\rm flavor}$. Since unlike the $\b$-deformed theory, the gauge field is neutral while
the gaugino is charged under the flavor symmetry, supersymmetry
is completely broken.

As in the $\b$-deformed case, also the $\gamma$-deformed action can be written
using  a noncommutative $*$-product between the fields
defined as
\eqn\star{
f*g=e^{i\pi\gamma_i\varepsilon^{ijk}Q^f_j Q^g_k}fg}
where $fg$ is the ordinary product and $Q_i,i=1,2,3$ are the $U(1)^3_{\rm flavor}$ charges of the fields.
Again,
by a direct inspection of Feynman diagrams
it is straightforward to see that the only modification of planar
diagrams compared to ${\cal N}=4$ SYM, is the multiplication by an overall phase, which depends only
on the flavor charges of the fields on the external lines. In particular if
all the fields on the external lines are gluons, the phase is 1 and
the scattering amplitudes are same as in the
undeformed   ${\cal N}=4$ SYM theory. This holds to all orders in perturbation theory for any
$n$-point function.
Note, that unlike the $\b$-deformed theory, here the gluino scattering amplitudes differ from those
of the  ${\cal N}=4$ SYM since they are charged under the flavor symmetry group.
Also, the same argument can be employed in the computation of the expectation value
of the Wilson loop in the planar limit of the $\gamma$-deformed theory and shows that it agrees with the ${\cal N}=4$ SYM result to all
orders in perturbation theory.

The dual gravity background
of the $\gamma$-deformed theory at large $N$
 was constructed in \FrolovDJ, using three TsT transformations. The metric
is of the type $AdS_5 \times {\cal M}_5$, where ${\cal M}_5$ is a deformation of $S^5$
\eqn\metrictwo{
ds^2_{{\cal M}_5} = \sum_{i=1}^3 (d\mu_i^2 + G \mu_i^2 d \phi_i^2) + R^2 G \mu_1^2\mu_2^2\mu_3^2(\sum_{i=1}^3 \gamma_i d\phi_i)^2
}
where $G^{-1} = 1 + R^2  (\gamma_3 \mu_1^2\mu_2^2+\gamma_1\mu_2^2\mu_3^2+\gamma_2\mu_1^2\mu_3^2)$ and $R$ is the $S^5$ radius.
Again, in addition to this deformation of the $S^5$-part of the metric and the
field strength $F^{(5)}$, the deformed nonsupersymmetric background has all
bosonic Type IIb supergravity fields turned on and affects only the compact part ${\cal M}_5$.
Thus, the T-duality transformation in the $AdS_5$ part, that is used in order to map
the scattering amplitude computation to a Wilson loop one is
valid also in the $\gamma$-deformed geometry.

A Green-Schwarz $\sigma$-model describing a closed string propagating
on the $\gamma$-deformed
background was constructed in \AAF.
Again, the  deformation of the
Green-Schwarz  $\sigma$-model action can be replaced by a modification of the periodicity
properties of the world-sheet fields.
The same arguments that we employed in the $\b$-deformed case can be used here:
the  Noether currents,
whose charge are responsible for the
non-periodicity of the fields depend only on $\p^\alpha \phi_i$ and vanish
for constant $\phi_i$, as is
the case in the internal part of the minimal area solution.
The
fermionic fields are still periodic
and the analysis of the spectrum of fluctuations is unmodified by the deformation.
Therefor, the corrections to any loop order
in the $AdS_5\times S^5$ and the $\gamma$-deformed $\sigma$-models are identical.


\vskip1cm
\medskip
{\bf Acknowledgment:} We acknowledge useful discussion with O. Aharony, S. Ananth, G. Arutyunov and T. McLoughlin.

\vskip1cm

\listrefs

\bye